\documentclass[a4paper,12pt,eqno]{article}
\usepackage{theorem}
\usepackage{latexsym,amssymb,amsfonts,amsmath}
\usepackage{graphicx}
\newtheorem{thm}{Theorem}[section]

\newtheorem{cor}[thm]{Corollary}
\newtheorem{pro}[thm]{Proposition}

\newtheorem{con}[thm]{Conjecture}

\theoremheaderfont{\scshape}

\title{{\Large {\bf On the relation between quantum walks \\ and zeta functions}}
\author{
{\small Norio Konno}\\
{\scriptsize Department of Applied Mathematics, 
Faculty of Engineering, 
Yokohama National University}\\
{\scriptsize Hodogaya, Yokohama 240-8501, Japan}\\
{\scriptsize e-mail: konno@ynu.ac.jp}\\
{\small Iwao Sato}\\
{\scriptsize Oyama National College of Technology}\\
{\scriptsize Oyama, Tochigi 323-0806, Japan}\\
{\scriptsize e-mail: isato@oyama-ct.ac.jp}\\}
}
\date{\empty }
\begin{document}
\maketitle

\par\noindent
\begin{small}
\par\noindent
{\bf Abstract}. We present an explicit formula for the characteristic polynomial of the transition matrix of the discrete-time quantum walk on a graph via the second weighted zeta function. As applications, we obtain new proofs for the results on spectra of the transition matrix and its positive support. 
  
\footnote[0]{
{\it Abbr. title:} On the relation between quantum walks and zeta functions
}
\end{small}

\setcounter{equation}{0}
\section{Introduction}
As a quantum counterpart of the classical random walk, the quantum walk has recently attracted much attention for various fields. The review and book on quantum walks are Ambainis \cite{Ambainis2003}, Kempe \cite{Kempe2003}, Kendon \cite{Kendon2007}, Konno \cite{Konno2008b}, for examples. Quantum walks of graphs were applied in graph isomorphism problems. 
Graph isomorphism problems determine whether two graphs are isomorphic. 
Shiau et al. \cite{ShiauETAL} first pointed out the deficiency of the simplest classical 
algorithm and continuous-time one particle quantum random walks 
in distinguishing some non-isomorphic graphs. 
Emms et al. \cite{EmmsETAL2009} introduced a graph-spectral technique induced 
by discrete-time quantum walks to distinguish two non-isomorphic graphs 
that are cospectral with respect to standard matrix representations. 
Gambel et al. \cite{GambelETAL} developed a method of characterizing the additional power 
that quantum walks of interacting particles have for distinguishing non-isomorphic regular graphs. 
Emms et al. \cite{EmmsETAL2006} treated spectra of the transition matrix and its positive support of 
the discrete-time quantum walk on a graph, and showed that the third power of the transition matrix 
outperforms the graph spectra methods in distinguishing strongly regular graphs. 
Godsil and Guo \cite{GG2010} gave new proofs of the results of Emms et al. \cite{EmmsETAL2006}.  

Already, the Ihara zeta function of a graph obtained various success related to graph spectra. 
Ihara zeta functions of graphs started from Ihara zeta functions of regular 
graphs by Ihara \cite{Ihara1966}. 
Originally, Ihara \cite{Ihara1966} presented $p$-adic Selberg zeta functions of 
discrete groups, and showed that its reciprocal is an explicit polynomial. 
Serre \cite{Serre} pointed out that the Ihara zeta function is the zeta function of 
the quotient $T/ \Gamma $ (a finite regular graph) of the one-dimensional 
Bruhat-Tits building $T$ (an infinite regular tree) associated with 
$GL(2, k_p)$. 
A zeta function of a regular graph $G$ associated with a unitary 
representation of the fundamental group of $G$ was developed by 
Sunada \cite{Sunada1986, Sunada1988}. 
Hashimoto \cite{Hashimoto1989} treated multivariable zeta functions of bipartite graphs. 
Bass \cite{Bass1992} generalized Ihara's result on the zeta function of 
a regular graph to an irregular graph, and showed that its reciprocal is 
again a polynomial. 
Various proofs of Bass' Theorem were given by 
Stark and Terras \cite{ST1996}, Foata and Zeilberger \cite{FZ1999}, Kotani and Sunada \cite{KS2000}. 
Sato \cite{Sato} defined a new zeta function of a graph by using not an infinite 
product but a determinant. 

Ren et al. \cite{RenETAL} found an interesting relationship between the Ihara zeta function and 
the discrete-time quantum walk on a graph, and showed that the support of 
the transition matrix of the discrete-time quantum walk is equal to 
the Perron-Frobenius operator (the edge matrix) related to the Ihara zeta function. Based on this analysis, Ren et al. explained that the Ihara zeta function 
can not distinguish cospectral regular graphs. 

It is known that a necessary condition of localization is the existence of degenerate eigenvalues of the transition matrix in the case of a three-state quantum walk (see Section 1.2 of \cite{Konno2008b}, for example). Our main result (Theorem {\rmfamily \ref{thm4.1}}) gives an explicit formula for the characteristic polynomial of the transition matrix of the quantum walk. So the result is useful for investigating dynamics of the walk and would be helpful to the study on transmission of the quantum information.
 
The rest of the paper is organized as follows. 
Section 2 gives the definition of the transition matrix of the discrete-time quantum walk on a graph, 
and review results on it. 
In Sect. 3, we define the Ihara zeta function and the second weighted zeta function of a graph, 
and present their determinant expressions. 
In Sect. 4, we present our main result (Theorem {\rmfamily \ref{thm4.1}}) of this paper. 
As a corollary, we give another proof of a result by Emms et al. \cite{EmmsETAL2006} on spectra of the transition matrix. 
In Sect. 5, we present another proof of a result by Emms et al. \cite{EmmsETAL2006} on spectra of the positive support of 
the transition matrix.

\section{Definition of the transition matrix of a quantum walk on a graph} 

Graphs treated here are finite. Let $G=(V(G),E(G))$ be a connected graph (possibly multiple edges and loops) 
with the set $V(G)$ of vertices and the set $E(G)$ of unoriented edges $uv$ 
joining two vertices $u$ and $v$. 
For $uv \in E(G)$, an arc $(u,v)$ is the oriented edge from $u$ to $v$. 
Set $D(G)= \{ (u,v),(v,u) | uv \in E(G) \} $. 
For $e=(u,v) \in D(G)$, set $u=o(e)$ and $v=t(e)$. 
Furthermore, let $e^{-1}=(v,u)$ be the {\em inverse} of $e=(u,v)$. 
The {\em degree} $\deg v = \deg {}_G \  v$ of a vertex $v$ of $G$ is the number of edges incident to $v$. 
For a natural number $k$, a graph $G$ is called {\em $k$-regular } if $\deg {}_G \  v=k$ for each vertex $v$ of $G$. 

A discrete-time quantum walk is a quantum analog of the classical random walk on a graph whose state vector is governed by 
a matrix called the transition matrix.  
Let $G$ be a connected graph with $n$ vertices and $m$ edges, 
$V(G)= \{ v_1 , \ldots , v_n \} $ and $D(G)= \{ e_1 , \ldots , e_m , 
e^{-1}_1 , \ldots , e^{-1}_m \} $. 
Set $d_j = d_{u_j} = \deg v_j $ for $j=1, \ldots ,n$. 
The {\em transition matrix} ${\bf U} ={\bf U} (G)=( U_{ef} )_{e,f \in D(G)} $ 
of $G$ is defined by 
\[
U_{ef} =\left\{
\begin{array}{ll}
2/d_{t(f)} (=2/d_{o(e)} ) & \mbox{if $t(f)=o(e)$ and $f \neq e^{-1} $, } \\
2/d_{t(f)} -1 & \mbox{if $f= e^{-1} $, } \\
0 & \mbox{otherwise.}
\end{array}
\right. 
\]

We introduce the {\em positive support} $\>{\bf F}^+ =( F^+_{ij} )$ of 
a real matrix ${\bf F} =( F_{ij} )$ as follows: 
\[
F^+_{ij} =\left\{
\begin{array}{ll}
1 & \mbox{if $F_{ij} >0$, } \\
0 & \mbox{otherwise.}
\end{array}
\right.
\]

Let $G$ be a connected graph. 
If the degree of each vertex of $G$ is not less than 2, i.e., $\delta (G) \geq 2$, 
then $G$ is called an {\em md2 graph}. 

The transition matrix of a discrete-time quantum walk in a graph 
is closely related to the Ihara zeta function of a graph. 
We stare a relationship between the discrete-time quantum walk and the Ihara zeta function of a graph by Ren et al. \cite{RenETAL}.

\begin{thm}[Ren, Aleksic, Emms, Wilson and Hancock \cite{RenETAL}] 
\label{thm2.1}
Let ${\bf B} - {\bf J}_0 $ be the Perron-Frobenius operator (or the edge matrix) 
of a simple graph subject to the md2 constraint, where the edge matrix is defined in Section 3. 
Let ${\bf U}$ be the transition matrix of the discrete-time quantum walk on $G$. 
Then the ${\bf B} - {\bf J}_0 $ is the positive support of the transpose of ${\bf U} $, i.e., 
\[
{\bf B} - {\bf J}_0 =( {}^T {\bf U} )^+, 
\]
where ${}^T {\bf U} $ is the transpose of ${\bf U}$. 
\end{thm}

\section{The Ihara zeta function of a graph}

Let $G$ be a connected graph. Then a {\em path $P$ of length $n$} in $G$ is a sequence $P=(e_1, \ldots ,e_n )$ of $n$ arcs such that $e_i \in D(G)$, $t( e_i )=o( e_{i+1} )(1 \leq i \leq n-1)$, where indices are treated $mod \  n$. 
Set $ | P | =n$, $o(P)=o( e_1 )$ and $t(P)=t( e_n )$. 
Also, $P$ is called an {\em $(o(P),t(P))$-path}. 
We say that a path $P=(e_1, \ldots ,e_n )$ has a {\em backtracking} 
if $ e^{-1}_{i+1} =e_i $ for some $i (1 \leq i \leq n-1)$. 
A $(v, w)$-path is called a {\em $v$-cycle} 
(or {\em $v$-closed path}) if $v=w$. 
The {\em inverse cycle} of a cycle 
$C=( e_1, \ldots ,e_n )$ is the cycle 
$C^{-1} =( e^{-1}_n , \ldots ,e^{-1}_1 )$.

We introduce an equivalence relation between cycles. 
Two cycles $C_1 =(e_1, \ldots ,e_m )$ and 
$C_2 =(f_1, \ldots ,f_m )$ are called {\em equivalent} if there exists 
$k$ such that $f_j =e_{j+k} $ for all $j$. 
The inverse cycle of $C$ is in general not equivalent to $C$. 
Let $[C]$ be the equivalence class which contains a cycle $C$. 
Let $B^r$ be the cycle obtained by going $r$ times around a cycle $B$. 
Such a cycle is called a {\em power} of $B$. 
A cycle $C$ is {\em reduced} if 
$C$ has no backtracking. 
Furthermore, a cycle $C$ is {\em prime} if it is not a power of 
a strictly smaller cycle. 
Note that each equivalence class of prime, reduced cycles of a graph $G$ 
corresponds to a unique conjugacy class of 
the fundamental group $ \pi {}_1 (G,v)$ of $G$ at a vertex $v$ of $G$. 

The {\em Ihara zeta function} of a graph $G$ is 
a function of $t \in {\bf C}$ with $| t |$ sufficiently small, 
defined by 
\[
{\bf Z} (G, t)= {\bf Z}_G (t)= \prod_{[C]} (1- t^{ | C | } )^{-1} ,
\]
where $[C]$ runs over all equivalence classes of prime, reduced cycles 
of $G$. 

Let $G$ be a connected graph with $n$ vertices and $m$ edges. 
Two $2m \times 2m$ matrices 
${\bf B} = {\bf B} (G)=( {\bf B}_{ef} )_{e,f \in D(G)} $ and 
${\bf J}_0 ={\bf J}_0 (G) =( {\bf J}_{ef} )_{e,f \in D(G)} $ 
are defined as follows: 
\[
{\bf B}_{ef} =\left\{
\begin{array}{ll}
1 & \mbox{if $t(e)=o(f)$, } \\
0 & \mbox{otherwise,}
\end{array}
\right.
 \qquad 
{\bf J}_{ef} =\left\{
\begin{array}{ll}
1 & \mbox{if $f= e^{-1} $, } \\
0 & \mbox{otherwise.}
\end{array}
\right.
\]
Then the matrix ${\bf B} - {\bf J}_0 $ is called the {\em edge matrix} of $G$.

\begin{thm}[Hashimoto \cite{Hashimoto1989}; Bass \cite{Bass1992}]
\label{thm3.1}
Let $G$ be a connected graph. Then the reciprocal of the Ihara zeta function of $G$ is given by 
\[
{\bf Z} (G, t)^{-1} =\det ( {\bf I}_{2m} -t ( {\bf B} - {\bf J}_0 ))
=(1- t^2 )^{r-1} \det ( {\bf I}_{n} -t {\bf A} (G)+ 
t^2 ({\bf D} -{\bf I}_{n})), 
\]
where ${\bf I}_{n}$ is the $n \times n$ identity matrix, $r$ and ${\bf A} (G)$ are the Betti number and the adjacency matrix 
of $G$, respectively, and ${\bf D} =( d_{ij} )$ is the diagonal matrix 
with $d_{ii} = \deg v_i $, $V(G)= \{ v_1 , \ldots , v_n \} $. 
\end{thm}

Let $G$ be a connected graph and $V(G)= \{ v_1 , \ldots , v_n \}$. 
Then we consider an $n \times n$ matrix 
${\bf W} =( w_{ij} )_{1 \leq i,j \leq n }$ with $ij$ entry 
nonzero complex number $w_{ij}$ if $( v_i , v_j ) \in D(G)$, 
and $w_{ij} =0$ otherwise. 
The matrix ${\bf W} = {\bf W} (G)$ is called the 
{\em weighted matrix} of $G$.
Furthermore, let $w( v_i , v_j )= w_{ij}, \  v_i , v_j \in V(G)$ and 
$w(e)= w_{ij}, e=( v_i , v_j ) \in D(G)$. 
For each path $P=( e_1 , \ldots , e_r )$ of $G$, the {\em norm} 
$w(P)$ of $P$ is defined as follows: 
$w(P)= w (e_1 ) w(e_2 ) \cdots w (e_r )$.

Let $G$ be a connected graph with $n$ vertices and $m$ edges, 
and ${\bf W} = {\bf W} (G)$ a weighted matrix of $G$.
A $2m \times 2m$ matrix 
${\bf B}_w = {\bf B}_w (G)=( {\bf B}^{(w)}_{ef} )_{e,f \in D(G)} $ is defined as follows: 
\[
{\bf B}^{(w)}_{ef} =\left\{
\begin{array}{ll}
w(f) & \mbox{if $t(e)=o(f)$, } \\
0 & \mbox{otherwise.}
\end{array}
\right.
\]
Then the {\em second weighted zeta function} of $G$ is defined by 
\[
{\bf Z}_1 (G,w,t)= \det ( {\bf I}_{2m} -t ( {\bf B}_w - {\bf J}_0 ) )^{-1} . 
\]
If $w(e)=1$ for any $e \in D(G)$, then the second weighted zeta function of $G$ is the 
Ihara zeta function of $G$.

\begin{thm}[Sato \cite{Sato}]
\label{thm3.2}
Let $G$ be a connected graph, and 
let ${\bf W} = {\bf W} (G)$ be a weighted matrix of $G$. 
Then the reciprocal of the second weighted zeta function of $G$ is given by 
\[
{\bf Z}_1 (G,w,t )^{-1} =(1- t^2 )^{m-n} 
\det ({\bf I}_n -t {\bf W} (G)+ t^2 ( {\bf D}_w - {\bf I}_n )) , 
\]
where $n = | V(G) | $, $m = | E(G) |$ and 
${\bf D}_w =( d_{ij} )$ is the diagonal matrix 
with $d_{ii} = \sum_{o(e)= v_i } w(e)$, $V(G)= \{ v_1 , \ldots , v_n \} $. 
\end{thm}

\section{The characteristic polynomial of the transition matrix}

We present a formula for the characteristic polynomial of ${\bf U}$. Let $G$ be a connected graph with $n$ vertices and $m$ edges. Then the $n \times n$ matrix ${\bf T} (G)=( T_{uv} )_{u,v \in V(G)}$ is given as follows: 
\[
T_{uv} =\left\{
\begin{array}{ll}
1/( \deg {}_G u)  & \mbox{if $(u,v) \in D(G)$, } \\
0 & \mbox{otherwise.}
\end{array}
\right.
\] 

\begin{thm}
\label{thm4.1}
Let $G$ be a connected graph with $n$ vertices $v_1 , \ldots , v_n $ and $m$ edges. Then, for the transition matrix ${\bf U}$ of $G$, we have
\begin{align*}
\det ( \lambda {\bf I}_{2m} - {\bf U} )
&= 
( \lambda {}^2 -1)^{m-n} \det (( \lambda {}^2 +1) {\bf I}_n -2 \lambda {\bf T} (G)) \\
& = \frac{( \lambda {}^2 -1)^{m-n} \det (( \lambda {}^2 +1) {\bf D} -2 \lambda {\bf A} (G))}
{d_{v_1} \cdots d_{v_n }}. 
\end{align*}
\end{thm}

\par\noindent
{\bf Proof}.  Let $G$ be a connected graph with $n$ vertices and $m$ edges, $V(G)= \{ v_1 , \ldots , v_n \}$ 
and $D(G)= \{ e_1 , \ldots , e_m , e^{-1}_1 , \ldots , e^{-1}_m \}$. 
Set $d_j = d_{v_j } = \deg v_j $ for each $j=1, \ldots ,n$. 
Then we consider a $2m \times 2m$ matrix 
${\bf B}_d =( B_{ef} )_{e,f \in D(G)}$ given by 
\[
B_{ef} =\left\{
\begin{array}{ll}
2/d_{o(f)} & \mbox{if $t(e)=o(f)$, } \\
0 & \mbox{otherwise.}
\end{array}
\right.
\]
By Theorem {\rmfamily \ref{thm3.2}}, we see
\[
\det ( {\bf I}_{2m} -t ( {\bf B}_d - {\bf J}_0 ))=(1- t^2 )^{m-n} 
\det ({\bf I}_n -t {\bf W}_d (G)+ t^2 ( {\bf D}_d - {\bf I}_n )) , 
\]
where ${\bf W}_d (G)=(w_{uv} )_{u,v \in V(G)} $ and ${\bf D}_d =( d_{uv} )_{u,v \in V(G)} $ are given as follows: 
\[
w_{uv} =\left\{
\begin{array}{ll}
2/d_u & \mbox{if $(u,v) \in D(G)$, } \\
0 & \mbox{otherwise,}
\end{array}
\right.
 \qquad 
d_{uv} =\left\{
\begin{array}{ll}
2 & \mbox{if $u=v$, } \\
0 & \mbox{otherwise.}
\end{array}
\right.
\]
Note that 
\[
d_j \times (2/d_j) = 2 \  (1 \leq j \leq n) . 
\]
Thus, 
\[ 
\det ( {\bf I}_{2m} -t ( {}^T {\bf B}_d - {}^T {\bf J}_0 ))=(1- t^2 )^{m-n} 
\det ({\bf I}_n -t {\bf W}_d (G)+ t^2 {\bf I}_n ). 
\]

But, we have 
\[
{}^T {\bf B}_d - {}^T {\bf J}_0 = {\bf U} \  and \  {\bf W}_d (G) = 2 {\bf T} (G) . 
\]
Therefore,  
\[
\det ( {\bf I}_{2m} -t {\bf U} )= (1- t^2 )^{m-n} 
\det ((1+ t^2 ) {\bf I}_n -2t {\bf T} (G)) . 
\]

Now, let $t=1/ \lambda $. Then we get 
\[ 
\det \left( {\bf I}_{2m} - \frac{1}{ \lambda } {\bf U} \right) = 
\left( 1- \frac{1}{ \lambda {}^2 } \right)^{m-n} 
\det \left( \left( 1+ \frac{1}{ \lambda {}^2 } \right) {\bf I}_n - \frac{2}{ \lambda } {\bf T} (G) \right) . 
\]
Thus, 
\[
\det ( \lambda {\bf I}_{2m} - {\bf U} )= 
( \lambda {}^2 -1)^{m-n} \det (( \lambda {}^2 +1) {\bf I}_n -2 \lambda {\bf T} (G)) . 
\]

Next, we have  
\[
{\bf T} (G)= {\bf D}^{-1} {\bf A} (G) . 
\]
Then it follows that 
\begin{align*}
\det ( \lambda {\bf I}_{2m} - {\bf U} ) 
&= ( \lambda {}^2 -1)^{m-n} \det (( \lambda {}^2 +1) {\bf I}_n -2 \lambda {\bf D}^{-1} {\bf A} (G)) 
\\
&= ( \lambda {}^2 -1)^{m-n} \det {\bf D}^{-1} \det (( \lambda {}^2 +1) {\bf D} -2 \lambda {\bf A} (G)). 
\end{align*}
Since $ \det {\bf D}^{-1}=1/( d_{v_1} \cdots d_{v_n } )$, 
\begin{align*}
\det ( \lambda {\bf I}_{2m} - {\bf U} ) = 
\frac{( \lambda {}^2 -1)^{m-n} \det (( \lambda {}^2 +1) {\bf D} -2 \lambda {\bf A} (G))}
{d_{v_1} \cdots d_{v_n }} . 
\end{align*}
$\Box$

We can express the spectra of the transition matrix ${\bf U}$ by means of those of 
${\bf T} (G)$ (see \cite{EmmsETAL2006}). 
Let $Spec ({\bf F})$ be the spectra of a square matrix ${\bf F}$ .

\begin{cor}[Emms, Hancock, Severini and Wilson \cite{EmmsETAL2006}] 
\label{cor4.2}
Let $G$ be a connected graph with $n$ vertices and $m$ edges. 
The transition matrix ${\bf U}$ has $2n$ eigenvalues of the form 
\[
\lambda = \lambda {}_T \pm i \sqrt{1- \lambda {}^2_T } , 
\]
where $\lambda {}_T $ is an eigenvalue of the matrix ${\bf T} (G)$. 
The remaining $2(m-n)$ eigenvalues of ${\bf U}$ are $\pm 1$ with equal multiplicities. 
\end{cor}

\par\noindent
{\bf Proof}. By Theorem {\rmfamily \ref{thm4.1}}, we have 
\[
\det ( \lambda {\bf I}_{2m} - {\bf U} )= ( \lambda {}^2 -1)^{m-n} 
\prod_{ \lambda {}_T  \in Spec ({\bf T} (G))} ( \lambda {}^2 +1 -2 \lambda {}_T \lambda ) . 
\]
Solving $\lambda {}^2 +1 -2 \lambda {}_T \lambda =0$, we obtain 
\[
\lambda = \lambda {}_T \pm i \sqrt{1- \lambda {}^2_T }.
\]
The result follows. 
$\Box$
 
Emms et al. \cite{EmmsETAL2006} determined the spectra of the transition matrix ${\bf U}$ by examining the elements of the transition matrix of a graph and 
using the properties of the eigenvector of a matrix.  
And now, we could explicitly obtain the spectra of the transition matrix ${\bf U}$ from its characteristic polynomial.

\section{The positive support of the transition matrix of a graph}

By Theorem {\rmfamily \ref{thm2.1}}, we express the spectra of the positive support ${\bf U}^+ $ of the transition matrix of a regular graph $G$ by means of those of the adjacency matrix ${\bf A} (G)$ of $G$ (see \cite{EmmsETAL2006}).

\begin{thm}[Emms, Hancock, Severini and Wilson \cite{EmmsETAL2006}]
\label{thm5.1}
Let $G$ be a connected $k$-regular graph with $n$ vertices and $m$ edges, and 
$\delta (G) \geq 2$. 
The positive support ${\bf U}^+ $ has $2n$ eigenvalues of the form 
\[
\lambda = \frac{\lambda {}_A }{2} \pm i \sqrt{k-1- \lambda {}^2_A /4} , 
\]
where $\lambda {}_A $ is an eigenvalue of the matrix ${\bf A} (G)$. 
The remaining $2(m-n)$ eigenvalues of ${\bf U}^+$ are $\pm 1$ with equal multiplicities. 
\end{thm}

\par\noindent
{\bf Proof}. Let $G$ be a connected graph with $n$ vertices and $m$ edges, $V(G)= \{ v_1 , \ldots , v_n \}$ and $D(G)= \{ e_1 , \ldots , e_m , e^{-1}_1 , \ldots , e^{-1}_m \}$. 
Since $G$ is $k$-regular, we have ${\bf D}= k {\bf I}_n $. 
By Theorems {\rmfamily \ref{thm2.1}} and {\rmfamily \ref{thm3.1}}, we obtain 
\begin{align*}
\det ( {\bf I}_{2m} -t {\bf U}^+ ) 
&= 
\det ( {\bf I}_{2m} -t ( {}^T {\bf B} - {}^T {\bf J}_0 )) = \det ( {\bf I}_{2m} -t ( {\bf B} - {\bf J}_0 ))
\\
&= (1- t^2 )^{m-n} \det \left({\bf I}_n -t {\bf A} (G)+ t^2 ( {\bf D} - {\bf I}_n ) \right)
\\
&= (1- t^2 )^{m-n} \det \left({\bf I}_n -t {\bf A} (G)+ t^2 (k-1) {\bf I}_n \right). 
\end{align*}
Now we put $t=1/ \lambda$. Then we have 
\[ 
\det \left( {\bf I}_{2m} - \frac{1}{ \lambda } {\bf U}^+ \right) 
= \left( 1- \frac{1}{ \lambda {}^2 } \right)^{m-n} 
\det \left( \left( 1+ \frac{k-1}{ \lambda {}^2 } \right) {\bf I}_n - \frac{1}{ \lambda } {\bf A} (G) \right). 
\]
Thus, 
\begin{align*}
\det \left( \lambda {\bf I}_{2m} - {\bf U}^+ \right)
&= 
( \lambda {}^2 -1)^{m-n} \det (( \lambda {}^2 + k-1) {\bf I}_n - \lambda {\bf A} (G)) 
\\
&=
( \lambda {}^2 -1)^{m-n} 
\prod_{ \lambda {}_A  \in Spec ({\bf A} (G))} ( \lambda {}^2 + k-1 - \lambda {}_A \lambda ). 
\end{align*}
Solving $\lambda {}^2 + k-1 - \lambda {}_A \lambda = 0$, we get 
\[
\lambda = \frac{ \lambda {}_A }{2} \pm i \sqrt{k - 1 - \lambda {}^2_A /4}. 
\]
The result follows. 
$\Box$ 

Godsil and Guo \cite{GG2010} presented a new proof of Theorem {\rmfamily \ref{thm5.1}} by using linear algebraic technique. 

From an argument in the first part of this proof, we easily obtain the following formula of the characteristic polynomial of ${\bf U}^+$:

\begin{pro}
\begin{align*}
\det \left( \lambda {\bf I}_{2m} - {\bf U}^+ \right) 
= ( \lambda {}^2 -1)^{m-n} \det \left( ( \lambda {}^2 -1) {\bf I}_n - \lambda {\bf A} (G) + {\bf D} \right). 
\end{align*}
\end{pro}

Emms et al. \cite{EmmsETAL2006} found the eigenvalues of the positive support $( {\bf U}^2 )^+$ of the second power ${\bf U}^2$ of the transition matrix ${\bf U} $ of a regular graph. Furthermore, Godsil and Guo \cite{GG2010} expressed $( {\bf U}^2 )^+ $ in terms of $ {\bf U}^+ $ by using linear algebraic technique, 
and presented another proof of the result of Emms et al. \cite{EmmsETAL2006}. Emms et al. \cite{EmmsETAL2006} determined a necessary and sufficient condition for any array of $( {\bf U}^3 )^+$ of a strongly regular graph to be equal to 1, and proposed the following conjecture:

\begin{con}[Emms, Hancock, Severini and Wilson \cite{EmmsETAL2006}] 
Let $G$ and $H$ be strongly regular graphs with the same set of parameters. 
Then $G \cong H$ if and only if $Spec ( ( {\bf U} (G)^3 )^+ )=Spec( ( {\bf U} (H)^3 )^+ )$. 
\end{con}

If we apply the linear algebraic technique of Godsil and Guo \cite{GG2010} to $( {\bf U}^3 )^+$, and approach the characteristic polynomial of $( {\bf U}^3 )^+$, then it might be likely to determine the spectra of $( {\bf U}^3 )^+$. In consequence, it might bring a progress toward a settlement of this conjecture. Related to this conjecture, to find an explicit formula for the characteristic polynomial of the positive support $( {\bf U}^n )^+$ for any $n$ would be one of the interesting future problems.

\
\par\noindent
{\bf Acknowledgment.} The first author was partially supported by the Grant-in-Aid for Scientific Research (C) of Japan Society for the Promotion of Science (Grant No. 21540118). The second author was partially supported by the Grant-in-Aid for Scientific Research (C) of Japan Society for the Promotion of Science (Grant No. 19540154).
\
\par

\begin{small}
\bibliographystyle{jplain}

\end{small}

\end{document}